# DesignCon 2014

# Efficient Circuit-Level Implementation of Knuth-Based Balanced and Nearly-Balanced Codes

Pamela A. Riggs, Mayo Clinic

Barry K. Gilbert, Mayo Clinic
gilbert.barry@mayo.edu, 507-284-4056

Erik S. Daniel, Mayo Clinic

## Abstract

Coding schemes are often used in high-speed processor-processor or processor-memory busses in digital systems. In particular, we have introduced (in a 2012 DesignCon paper) a zero sum (ZS) signaling method which uses balanced or nearly-balanced coding to reduce simultaneous switching noise (SSN) in a single-ended bus to a level comparable to that of differential signaling. While several balanced coding schemes are known, few papers exist that describe the necessary digital hardware implementations of (known) balanced coding schemes, and no algorithms had previously been developed for nearly-balanced coding. In this work, we extend a known balanced coding scheme to accommodate nearly-balanced coding and demonstrate a range of coding and decoding circuits through synthesis in 65 nm CMOS. These hardware implementations have minimal impact on the energy efficiency and area when compared to current serializer/deserializers (SerDes) at clock rates which would support SerDes integration.

## Author Biographies

**Pamela A. Riggs** received a B.S. Degree in Electrical Engineering from Arizona State University in Tempe, AZ. She is currently a Project Engineer at the Mayo Clinic Special Purpose Processor Development Group. She designs high-speed digital integrated circuits.

**Barry K. Gilbert** received a BSEE from Purdue University (West Lafayette, IN) and a Ph.D. in physiology and biophysics from the University of Minnesota (Minneapolis, MN). He is currently Director of the Special Purpose Processor Development Group, directing research efforts in high performance electronics and related areas.

**Erik S. Daniel** received a BA in physics and mathematics from Rice University (Houston, TX) in 1992 and a Ph.D. degree in solid state physics from the California Institute of Technology (Pasadena, CA) in 1997. He is currently Deputy Director of the Special Purpose Processor Development Group, directing research efforts in high performance electronics and related areas.

# Introduction

For decades, a myriad of balanced coding schemes (i.e., those producing codewords with equal numbers of ones and zeroes) have been developed and reported (consider the representative subset [1]-[6]), primarily for error correction coding in storage, digital communications, or related applications. More recently, we have described an efficient high data rate parallel bus communications scheme capable of mitigating SSN which can be enabled by efficient balanced codes, or nearly-balanced codes (i.e. those producing codewords with a bounded difference between the number of ones and zeroes) [7]. Although there is significant literature describing scores of balanced coding schemes, few of these reports discuss efficient hardware implementations. In particular, in order to consider the incorporation of codes into digital systems such as high performance computers, it is necessary to quantify the integrated circuit resources required (e.g., chip area and/or number of gates), power consumed, and maximum clock rate. In this paper, we explore implementation of Knuth's original balanced coding methods [2] to enable balanced encoding, through HDL definition and subsequent synthesis into a 65 nm CMOS circuit technology in order to quantify speed, size, and power consumption metrics. We also demonstrate a simple method for extending the Knuth algorithms to accommodate nearly-balanced codes, and assess digital implementations of these algorithms as well.

The next section provides an overview of the Knuth balanced coding schemes (and nearly-balanced extensions thereof) employed for this work. The following section describes the circuit design details. The subsequent sections summarize the power, speed, and area of circuits synthesized across a range of design parameters.

# Knuth-Based Algorithms for Balanced and Nearly-Balanced Encoding / Decoding

Although there are many published articles discussing algorithms for balanced coding (as mentioned above), many/most of these algorithms are derived from a set of methods described by Knuth [2]. We found the original Knuth methods to be fairly simply described, and readily implemented in VHDL; hence they were chosen for the basis of this work. Furthermore, we found that we could fairly simply extend these Knuth methods to accommodate any finite amount of disparity, as is described below. Note that throughout, we will use the following notation: *n* = number of data bits, *p* = number of parity bits, and *m* = number of codeword bits (*m* = *n* + *p*). Further, we will use the following mathematical operators:

- $\lfloor x \rfloor$ is defined as the largest integer which is less than or equal to x
- $\lceil x \rceil$ is defined as the smallest integer which is greater than or equal to x
- $\binom{x}{y} \equiv \frac{x!}{y!(x-y)!}$

## Original Knuth Methods for Balanced Coding

In his original paper [2], Knuth described a number of balanced coding schemes, all based on the same basic idea. In this paper, we utilize two methods from the original paper, which Knuth referred to as the Simple Parallel and Optimized Parallel methods. We will refer to them as Knuth SP and Knuth OP, or simply SP and OP. Note that here we will use slightly different terminology than is used in Knuth's original paper, but the algorithm as described is mathematically identical. For simplicity, we restrict discussion to words of even length.

Knuth's Simple Parallel method proceeds as follows. A given input data word, $w$ of length $n$ is represented as a string of ones and zeroes (e.g., $w$=100101011101). We define $v(w)$ as the number of ones in $w$ minus the number of zeroes in $w$, such that a word $w$ is balanced if $v(w) = 0$. Note that through the course of this paper, we will refer to $v(w)$ as the "disparity" of the word $w$. We define $w^{(k)}$ to be the word $w$ with its first $k$ bits complemented. For example, if $w$=00110011, then $w^{(4)}$ = 11000011.

**Knuth Procedure Pseudo-Code**

- **Let w = n-bit input word**
- **For k = 0 to n-1**
  - **Compute $w^{(k)}$ = w with first k bits flipped**
  - **If $w^{(k)}$ has the same number of "1"s and "0"s, exit loop**
- **Next k**
- **Select balanced parity word $u_k$ which corresponds to number of bits flipped (k) in balanced word**
- **Output codeword = $u_k w^{(k)}$**

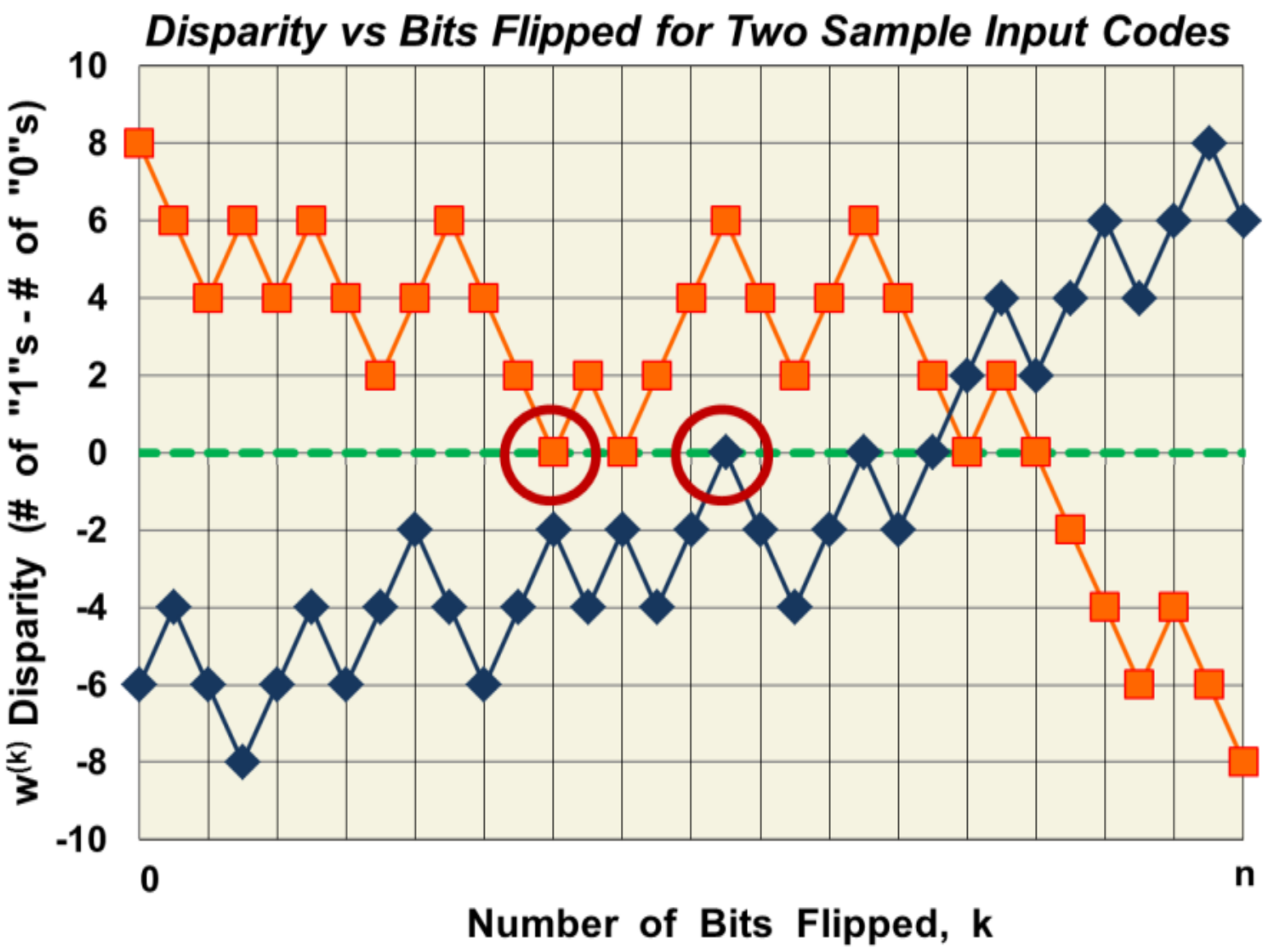

**Why Does This Work?**

- **Disparity at k=0 is the opposite sign of the disparity at k=n**
- **Every step in k results in a disparity change of +-2**
- **Therefore, the "random walk" associated with any input code must result in balance at some number of flipped bits**

**Figure 1: General Description of Knuth Simple Parallel Balanced Coding Algorithm (42515).**

Now, consider a word $w$ of length $n$, and consider the set of all $w^{(k)}$ as $k$ runs from 0 (no bits complemented) to $n$ (all bits complemented). We refer to this increment in $k$ (and corresponding bit flip) as one "step". Note that $v(w^{(0)}) = - v(w^{(n)})$. Also, note that as $k$ is incremented by 1 (and hence one additional bit is complemented), $v(w^{(k)})$ changes by ±2. Given these two facts, it must be true that for some $0 \leq k < n$, $v(w^{(k)}) = 0$, and hence the word $w^{(k)}$ is balanced. This algorithm is illustrated in Figure 1. A balanced codeword of length $p + n$ can be used to represent $w$ as follows. Encode the integer $k$ into a balanced word $u$ of (fixed) length $p$ using a simple table lookup. Then create the balanced codeword $uw^{(k)}$. This codeword can be decoded by using a table lookup to determine $k$ from $u$, then inverting the first $k$ bits of $w^{(k)}$ to recover $w$. An 8-bit example is illustrated in Figure 2.

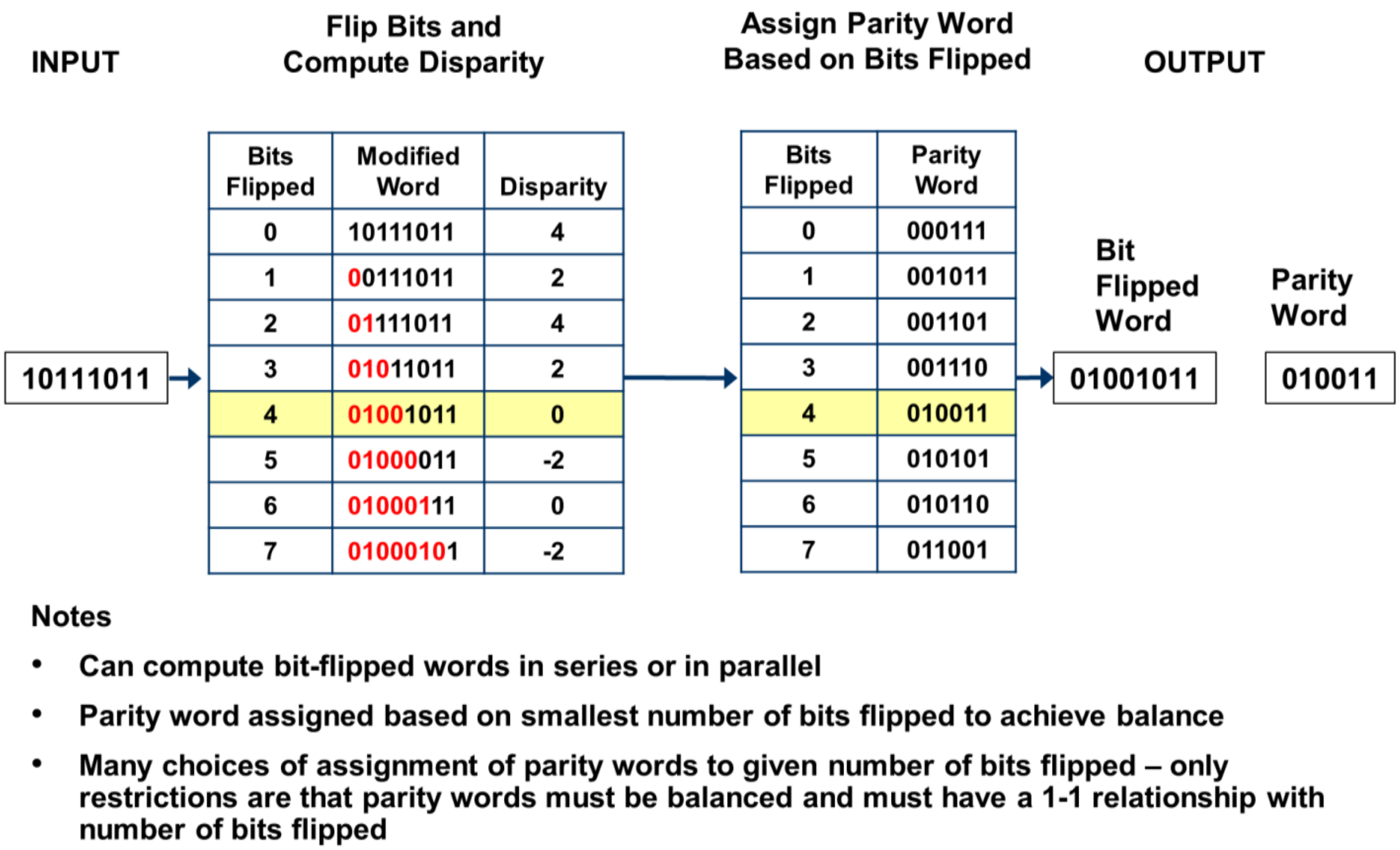

**Figure 2: Example of Knuth Simple Parallel Balanced Coding Algorithm (8-Bit Data Word) (42516).**

This Simple Parallel method is fairly readily implemented, but has one main shortcoming. The number of parity bits required, $p$, is greater than the theoretical minimum for balanced coding. The theoretical minimum number of parity bits, $p_{min}$, such that all $n$-bit data words can be represented by codewords of length $m_{min} = n + p_{min}$ must satisfy $\log_2 \binom{m_{min}}{\lfloor m_{min}/2 \rfloor} \geq n$ . Note that below (in Table 1), we will refer to this minimum codeword length as "Ideal ZS±0", referring to the ideal "zero sum" codeword length with disparity = 0. Although not shown in detail here, we can more generally compute the minimum codeword length for a given disparity, d, referred to as "Ideal ZS±d". Knuth demonstrated that the Simple Parallel method can be further improved (though still generally falling short of reaching an optimum number of parity bits) by allowing for parity words $\{u\}$ which are not necessarily balanced, making some adjustments to the algorithm accordingly. Knuth's Optimized Parallel method is one such algorithm, illustrated in Figure 3, and described here.

**Knuth Procedure Pseudo-Code**

- Let w = n-bit input word
- Select set of $\{k_j, u_j\}$ (see below)
- For j = 0 to n-1 + # of transitions
  - Compute $w^{(kj)}$ = w with first $k_j$ bits flipped
  - Select parity word $u_j$ which corresponds to number of bits flipped ($k_j$) in balanced word
  - If $u_j w^{(kj)}$ has the same number of "1"s and "0"s, exit loop
- Next j
- Output codeword = $u_j w^{(kj)}$

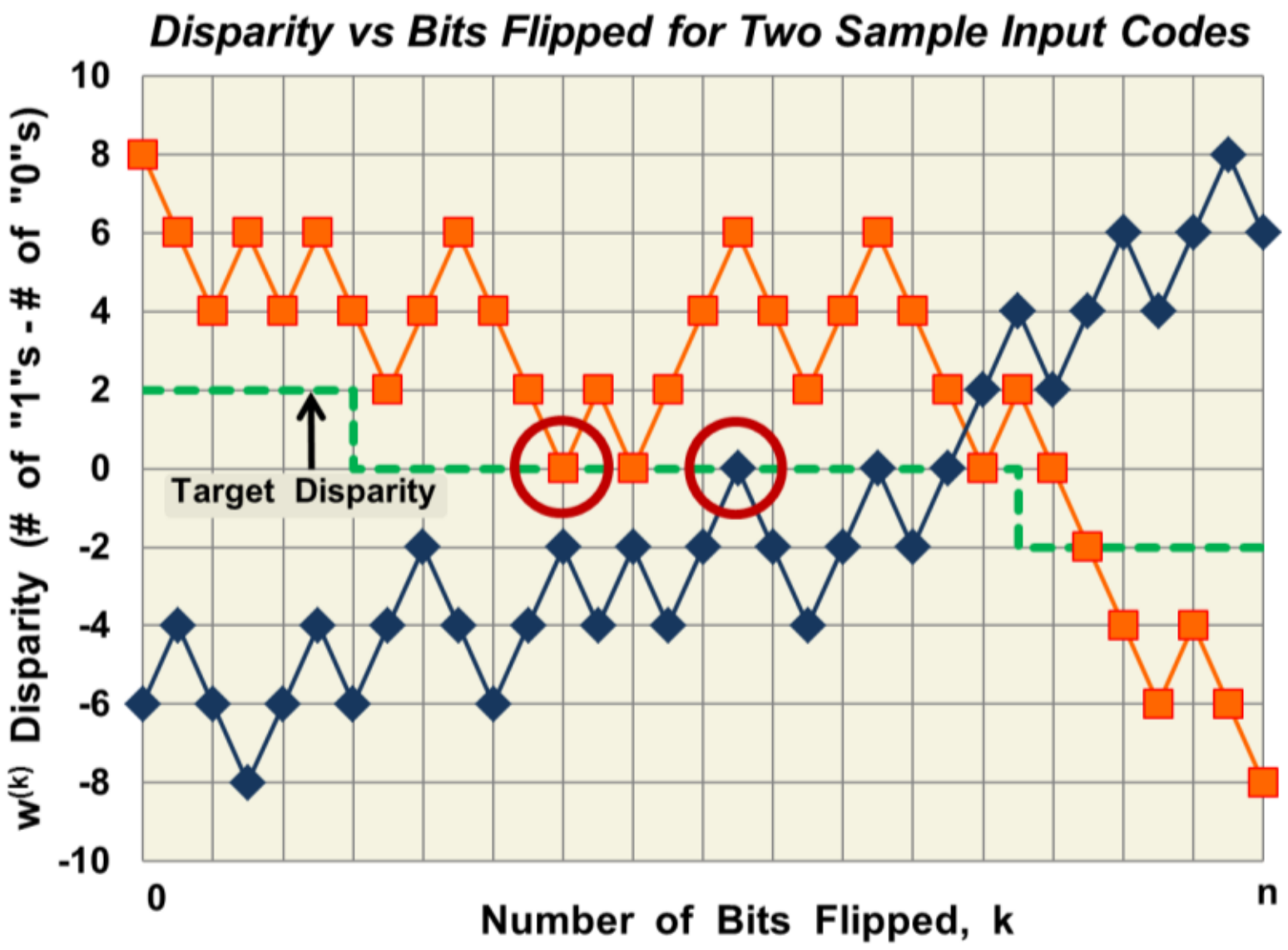

**Why Does This Work?**

- Same basic procedure as for Simple Parallel Algorithm, but $u_k$ not required to be balanced
  - Arrange $u_k$ in order from smallest to largest number of "1"s – corresponds to changing "target" number of "1"s in bit-flipped word
  - For each step in j, either increment k or number of ones in $u_j$, but not both
- Overall codeword disparity changes by +-2 with each step in j → must be balanced at some step (same basic argument applies here as for Simple Parallel Algorithm)

MAR_29 / 2011 / ESD / 42521
MAYO CLINIC
SPPDG

**Figure 3: General Description of Knuth Optimized Parallel Balanced Coding Algorithm (42521).**

Again, consider a word *w* of length *n*. Construct a set of pairs, $\{k_j, u_j\}$, where $k_j$ is the number of bits flipped in the data word at the *j*th step, and $u_j$ is the parity code corresponding to the *j*th step. These pairs are constructed to have the following properties: $v(u_{j+1}) - v(u_j) = 0$ or 1, and $k_{j+1} - k_j = 1 - (v(u_{j+1}) - v(u_j))$. In other words, as *j* is increased, either the number of bits flipped, $k_j$, increases, or the number of ones in the codeword $u_j$ increases (but not both). Note that the "target" disparity for the bit flipped word $w^{k_j}$ decreases every time $v(u_j)$ increases such that the overall codeword will be balanced. Knuth demonstrates that one can always construct such a set of pairs such that, using a very similar argument as for the Simple Parallel algorithm, a balanced codeword will be found for some *j*. As Knuth discusses [2], some care must be taken in the selection of the range of disparity of the parity words; we will not elaborate on this here. As the codewords, $u_j$, are no longer required to be balanced, this algorithm is somewhat more efficient than the Knuth Simple Parallel algorithm, generally requiring fewer parity bits. An 8-bit example is illustrated in Figure 4. In this example, only four parity bits are required, versus six in the Knuth Simple Parallel algorithm described above.

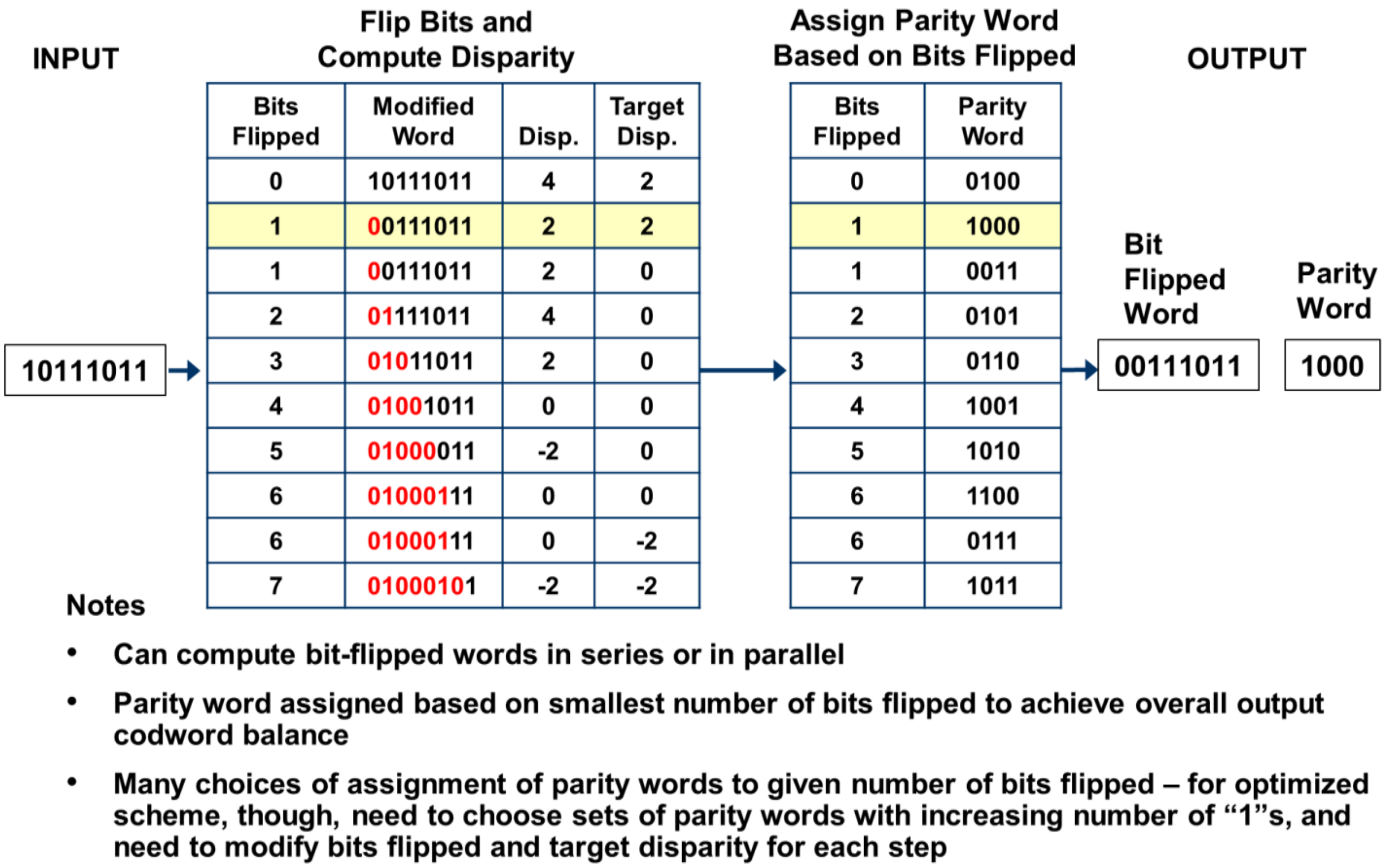

**Figure 4: Example of Knuth Optimized Parallel Balanced Coding Algorithm (8-Bit Data Word) (42522).**

## Nearly Balanced Extensions of Knuth Methods

As was discussed in the introduction of this paper, we wished to be able to assess coding schemes producing nearly-balanced codewords (i.e., those of low disparity such as ±2 or ±4) in addition to perfectly balanced codewords (i.e., disparity = 0). The advantage of nearly balanced codes is that they provide more available codewords, and therefore require fewer parity bits than perfectly balanced codes as shown in Table 1. Moreover, as shown in our original study of the effect of balanced or nearly balanced coding in suppressing SSN, codes with modest disparity were found to be nearly as effective as balanced codes in SSN suppression [7]. Unfortunately, we were unable to find any simple implementations of nearly balanced coding in the published literature. However, we found that the Knuth SP and OP methods could both be very simply extended to accommodate finite disparity coding. In both cases, these extensions rely on the fact that the disparity of the codeword changes by ±2 at each step in $k$ (in the case of the Simple Parallel method) or $j$ (in the case of the Optimized Parallel Method). Therefore, if finite disparity is allowed, one can eliminate a subset of these steps and still find an acceptable solution.

Consider extension of the SP method to allow for bounded disparity, $|v(w)| \leq 2d$ where $d$ is a positive integer. This method is illustrated for the case of $d$=1 (disparity ≤ ±2) in Figure 5, and in the case of $d$=2 (disparity ≤ ±4) in Figure 6. The standard SP procedure is followed, but instead of considering the set of $v(w^{(k)})$ for all $k$ from 0 to $n$, consider a subset of this set in which select elements are skipped. The minimum number of members in the set of $v(w^{(k)})$ needed (and hence the minimum number of $u$ words needed) is $N_u = \lceil n/(2d+1) \rceil$. If these are equally spaced, this places one member of the set no fewer than $d$ steps away from any of the members of the set which have been excluded, hence the disparity is bounded at $\pm 2d$. We can construct such a set by setting $k_j = \left\lfloor \left(\frac{(2j+1)(n-1)}{N_u}\right) + 0.5 \right\rfloor$ and considering the set of $v(w^{(kj)})$ for $j$ = 0 to $N_u$-1. Again, the integer $k_j$ can be represented by a balanced codeword, $u$ of length $p$ bits, and the codeword $uw^{(kj)}$ can be constructed. Note that for finite allowed disparity $2d$, the number of values of $k$ (i.e., $N_u$) is reduced by a factor of $2d$+1, and hence the necessary number of parity bits, $p$, is reduced by $\log_2(2d+1)$, thereby creating more efficient codes.

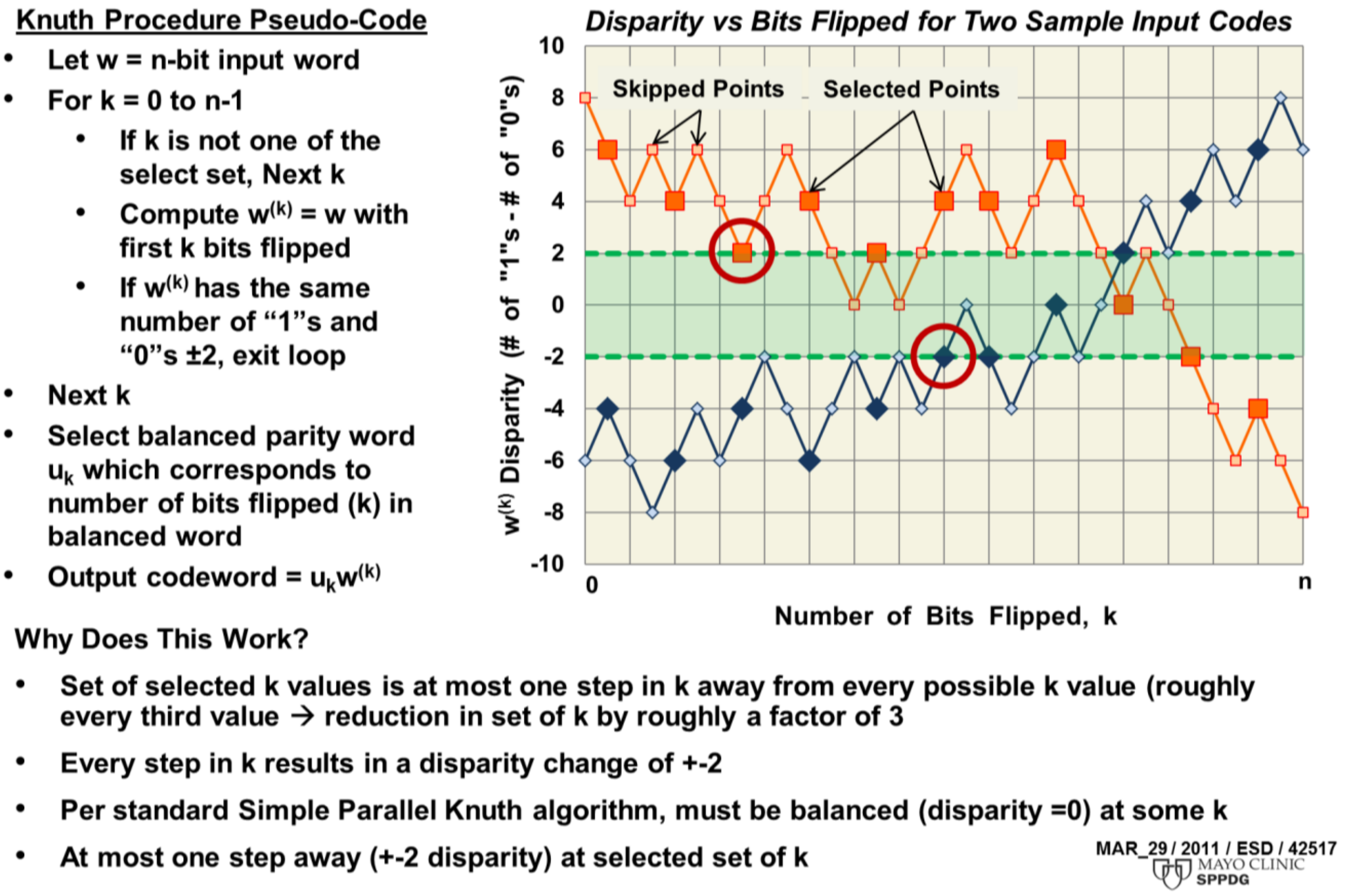

**Figure 5: General Description of Extension to Knuth Simple Parallel Balanced Coding Algorithm, Accommodating Finite disparity (±2 Disparity Example) (42517).**

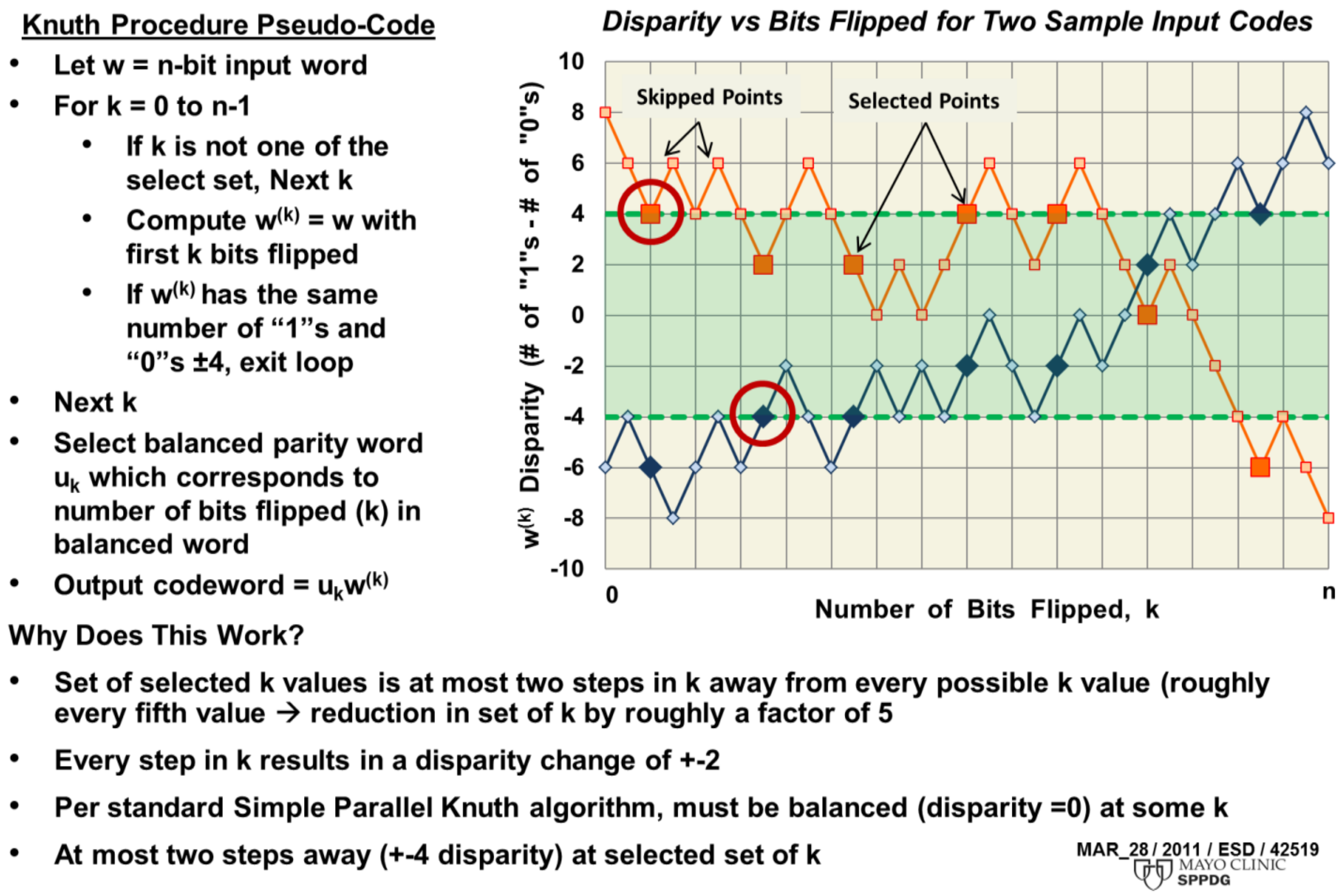

**Figure 6: General Description of Extension to Knuth Simple Parallel Balanced Coding Algorithm, Accommodating Finite Disparity (±4 Disparity Example) (42519).**

Note that this extended algorithm retains the simplicity of the original algorithm, and as will be demonstrated below, the extended algorithm is actually more efficient to implement in hardware or software as *d* increases. Examples for 8-bit words are shown in Figure 7 (d=1) and Figure 8 (d=2) respectively. Note that in the finite disparity cases the number of parity bits, *p*, is less than for the perfectly balanced case (6 bits for *d*=0, 4 bits for *d*=1, and 2 bits for *d*=2).

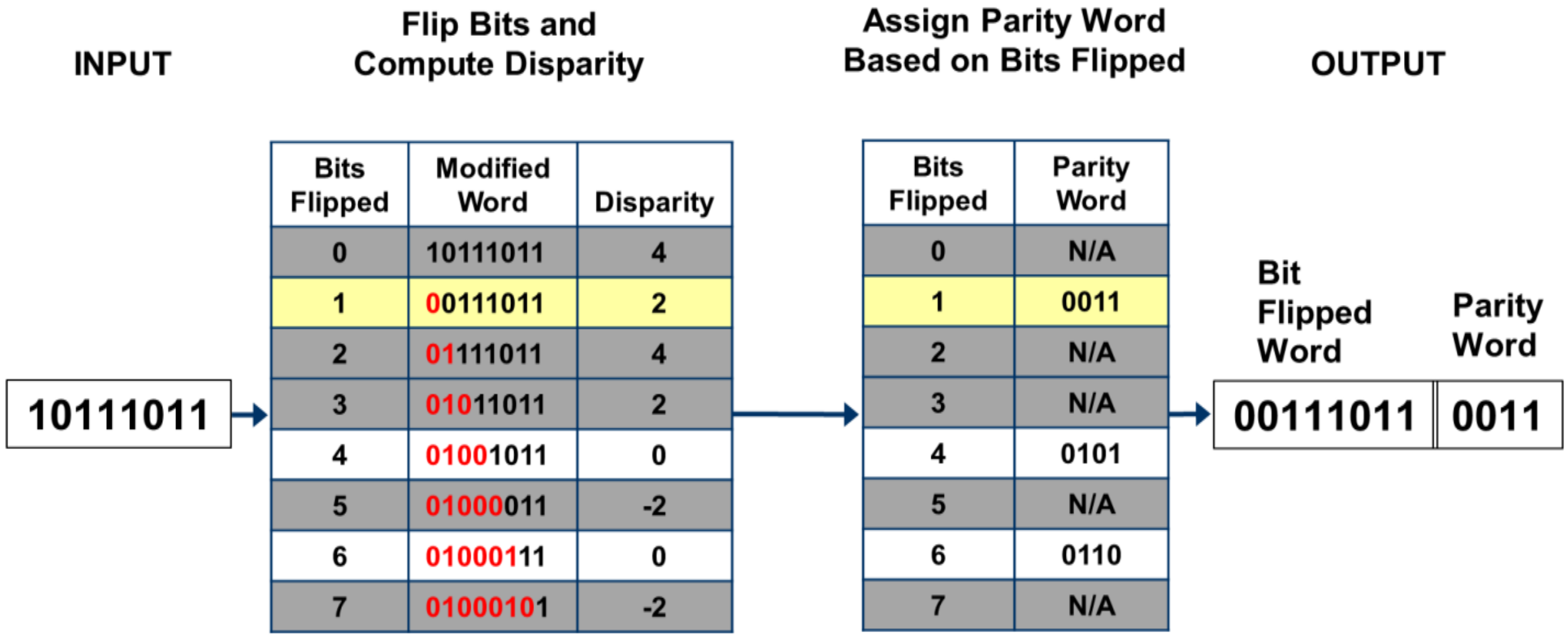

| Bits Flipped | Modified Word | Disparity |
|---|---|---|
| 0 | 10111011 | 4 |
| 1 | 00111011 | 2 |
| 2 | 01111011 | 4 |
| 3 | 01011011 | 2 |
| 4 | 01001011 | 0 |
| 5 | 01000011 | -2 |
| 6 | 01000111 | 0 |
| 7 | 01000101 | -2 |

| Bits Flipped | Parity Word |
|---|---|
| 0 | N/A |
| 1 | 0011 |
| 2 | N/A |
| 3 | N/A |
| 4 | 0101 |
| 5 | N/A |
| 6 | 0110 |
| 7 | N/A |

**Notes**

- Can compute bit-flipped words in series or in parallel
- Reduction in number of bit-flipped choices that need to be checked when compared with zero disparity case
  - Results in reduction in number of bits in parity word
  - Results in reduction in amount of logic needed

**Figure 7: Example of Extension to Knuth Simple Parallel Balanced Coding Algorithm Allowing Finite Disparity (8-Bit Data Word, ±2 Disparity) (42518).**

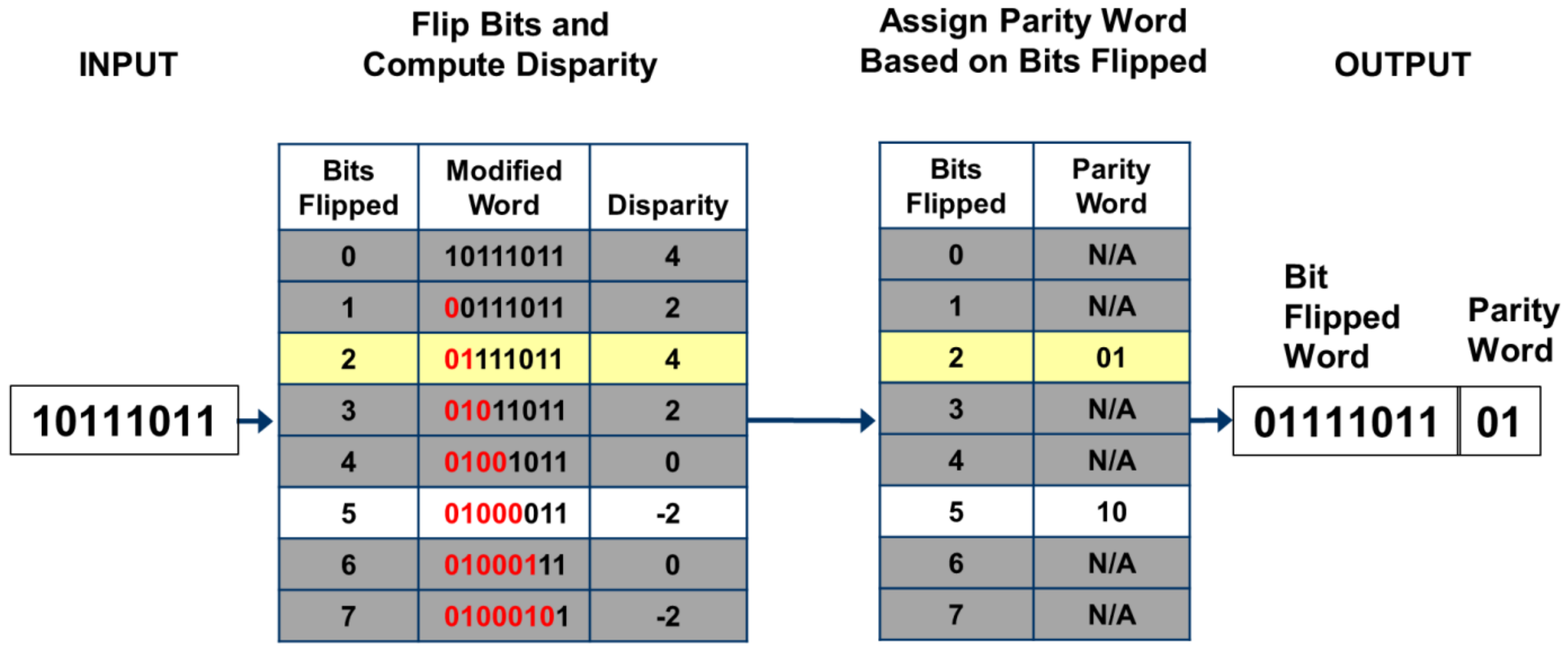

| Bits Flipped | Modified Word | Disparity |
|---|---|---|
| 0 | 10111011 | 4 |
| 1 | 00111011 | 2 |
| 2 | 01111011 | 4 |
| 3 | 01011011 | 2 |
| 4 | 01001011 | 0 |
| 5 | 01000011 | -2 |
| 6 | 01000111 | 0 |
| 7 | 01000101 | -2 |

| Bits Flipped | Parity Word |
|---|---|
| 0 | N/A |
| 1 | N/A |
| 2 | 01 |
| 3 | N/A |
| 4 | N/A |
| 5 | 10 |
| 6 | N/A |
| 7 | N/A |

**Notes**

- Can compute bit-flipped words in series or in parallel
- Reduction in number of bit-flipped choices that need to be checked when compared with zero disparity or ±2 disparity cases
  - Results in reduction in number of bits in parity word
  - Results in reduction in amount of logic needed

MAR_29 / 2011 / ESD / 42520
MAYO CLINIC
SPPDG

**Figure 8: Example of Extension to Knuth Simple Parallel Balanced Coding Algorithm Allowing Finite Disparity (8-Bit Data Word, ±4 Disparity) (42520).**

Although we do not discuss the details here, we note that the Knuth Optimized Parallel algorithm can be similarly extended.

In Table 1, we summarize the number of total bits required to "encode" a given number of data bits (*n*) across the entire range of encoding / signaling schemes described thus far in this section. In this table, "SE" refers to traditional single-ended signaling and "Diff" refers to traditional differential signaling. The "Ideal ZS" columns reflect the minimum (even) number of bits that would be required to provide at least $2^n$ codewords for the given level of allowed disparity (±0, ±2, or ±4), without identification of any specific encoding / decoding algorithm. The "SP" and "OP" columns refer to the Knuth Simple Parallel method and Knuth Optimized Parallel methods respectively, with or without extensions to accommodate finite disparity. As can be seen from this chart, the OP algorithm equals or improves upon the efficiency of the SP algorithm at all *n*. Also, as was mentioned above, increasing allowed disparity results in equal or greater efficiency of both the SP or OP algorithms. However, the SP and OP algorithms are not maximally efficient in most cases (i.e., total number of bits equal to the minimum number of bits listed in the corresponding "Ideal_ZS" column).

| **# of Data Bits, *n*** | **SE** | **Diff** | **Ideal ZS±0** | **Ideal ZS±2** | **Ideal ZS±4** | **SP±0** | **SP±2** | **SP±4** | **OP±0** | **OP±2** | **OP±4** |
|---|---|---|---|---|---|---|---|---|---|---|---|
| 4 | 4 | 8 | 6 | 6 | 4 | 8 | 6 | 6 | 8 | 6 | 6 |
| 6 | 6 | 12 | 8 | 8 | 8 | 10 | 8 | 8 | 10 | 8 | 8 |
| 8 | 8 | 16 | 12 | 10 | 10 | 14 | 12 | 10 | 12 | 10 | 10 |
| 10 | 10 | 20 | 14 | 12 | 12 | 16 | 14 | 12 | 14 | 12 | 12 |
| 12 | 12 | 24 | 16 | 14 | 14 | 18 | 16 | 16 | 16 | 16 | 14 |
| 14 | 14 | 28 | 18 | 16 | 16 | 20 | 18 | 18 | 20 | 18 | 16 |
| 16 | 16 | 32 | 20 | 18 | 18 | 22 | 20 | 20 | 22 | 20 | 18 |
| 20 | 20 | 40 | 24 | 22 | 22 | 26 | 26 | 24 | 26 | 24 | 24 |
| 24 | 24 | 48 | 28 | 26 | 26 | 32 | 30 | 28 | 30 | 28 | 28 |
| 28 | 28 | 56 | 32 | 30 | 30 | 36 | 34 | 32 | 34 | 32 | 32 |
| 32 | 32 | 64 | 36 | 34 | 34 | 40 | 38 | 38 | 38 | 36 | 36 |
| 40 | 40 | 80 | 44 | 42 | 42 | 48 | 46 | 46 | 46 | 44 | 44 |
| 48 | 48 | 96 | 52 | 50 | 50 | 56 | 54 | 54 | 54 | 54 | 52 |
| 64 | 64 | 128 | 68 | 66 | 66 | 72 | 72 | 70 | 72 | 70 | 68 |
| 72 | 72 | 144 | 76 | 74 | 74 | 82 | 80 | 78 | 80 | 78 | 76 |

**Table 1: Magnetic Number of Coded Bits Needed for Different Encoding / Signaling Schemes (Assuming an Even Number of Data and Parity Bits).**

## Circuit Implementation of Balanced and Nearly-Balanced Encoding / Decoding Blocks

In this section, we discuss translation of a number of Knuth and Knuth extension algorithms into realizable circuits in order to assess possible performance metrics (i.e. speed, power consumption, area required, and energy efficiency). In order to do so, first, several of the coding and decoding algorithms were translated into behavioral VHDL code. Variations of the code were written to accommodate changes in the following parameters: number of input bits, algorithm, architecture, and disparity.

A number of different design parameters were explored, summarized in Table 2.

| **Parameter Names** | **Description** |
|---|---|
| Number of Input Bits | Number of encoder input data bits which may be 8, 16, 32, or 64. |
| Algorithm | Either the Knuth simple parallel (SP) algorithm in which ones and zeroes in the parity word are balanced, or the Knuth Optimized Parallel (OP) coding algorithm, in which ones and zeroes in the parity word need not be balanced. |
| Architecture | Either Parallel (in which balance checkers work simultaneously (in parallel) to check the balance of inputs) or Pipeline (in which the balance checkers work in series). |
| Disparity | The number of ones minus the number of zeroes in the encoder output and decoder input. Values can be ±0, ±2, and ±4. |

**Table 2: Descriptions of Encoder and Decoder Design Parameters.**

Later discussions will refer to the designs using parameters from Table 2. The labels will be of the form: “<algorithm> <architecture> <disparity>”. The number of data bits or *n*, as described in the “Knuth-Based Algorithms for Balanced and Nearly-Balanced Encoding / Decoding” section, may or may not be mentioned. The decoders have the same architecture so that description will not be in the label. An example description is 64-bit “SP Pipeline ±2” which is an encoder with a 64-bit input, employing the Knuth SP coding scheme, built as a pipelined architecture, and a disparity of ±2. Note that the inclusion of an architecture type in the label means that the description is for an encoder.

The methodologies used to develop and analyze the circuit, the design variations, and the design architectures are described in the following portions of this section.

## Circuit Development and Analysis Methodologies

Synthesis into the target 65 nm CMOS ASIC library was conducted with Cadence’s RTL Compiler software to assess performance metrics. Unit time simulations verified functionality. Place and route were not performed, but synthesized designs were required to operate at the target frequency with at least 100 ps slack in order to allow for delays associated with routing. As all designs implemented were relatively small (at most ~570 µm × 570 µm, corresponding to ~4 ps transit time across the entire circuit), we believe a 100 ps slack requirement is a conservative constraint.

We note that the results presented in this study are relatively conservative, in that many additional optimizations could be applied.

## Circuit Configurations

Four parameters of the zero sum encoders were varied for this study. First, *n* was varied from 8-bits to 64-bits. Second, the designs employed the Knuth SP or OP algorithms. Third, the circuit architecture employed to realize the encoding operation was a parallel or pipeline architecture and is discussed in further detail in the "Parallel and Pipelined Encoder Architectures" section. Note that in descriptions of encoders the use of the word "parallel" refers to the architecture of the design and not the type of Knuth algorithm. Fourth, the disparity, discussed in more detail in the "Knuth-Based Algorithms for Balanced and Nearly-Balanced Encoding / Decoding" section, was varied from ±0, which is a balanced design, to ±2 and ±4, which are nearly-balanced designs. Given the large number of possible combinations, we did not attempt to evaluate every possible combination of these parameters. Rather, we chose what we felt to be a representative sampling of the possible parameter space.

The VHDL code for the decoders was simple enough that only one type of architecture was used. Therefore, only *n*, the type of algorithm, and the disparity were varied for each decoder.

Table 3 lists the encoder and decoder configurations that were synthesized and simulated for this study. It should be noted that the 16-bit OP Parallel ±2 and 16-bit SP Parallel ±2 encoders are the same design and the same applies to their corresponding decoders. Table 1 shows that for $n$=16 the number of coded bits needed for the two designs is the same, so in this particular case, the OP algorithm reduces to the same design as the SP algorithm: all parity words are balanced.

| No. of Data Bits | Design Type | Knuth Algorithm Type | Architecture | Disparity |
|---|---|---|---|---|
| 8, 16, 32, 64 | Encoder | SP | Parallel | ±0 |
| 8, 16, 32, 64 | Encoder | SP | Parallel | ±2 |
| 8, 16, 32, 64 | Encoder | SP | Parallel | ±4 |
| 8, 16, 32, 64 | Encoder | OP | Parallel | ±2 |
| 8, 16, 32, 64 | Encoder | SP | Pipeline | ±2 |
| 8, 16, 32, 64 | Decoder | SP | N/A | ±0 |
| 8, 16, 32, 64 | Decoder | SP | N/A | ±2 |
| 8, 16, 32, 64 | Decoder | SP | N/A | ±4 |
| 8, 16, 32, 64 | Decoder | OP | N/A | ±2 |

**Table 3: Encoder and Decoder Configurations Evaluated.**

## Parallel and Pipelined Encoder Architectures

Two architectures were considered for the encoder: parallel and pipelined. Both architectures invert bits in the input data until it is balanced or nearly-balanced then output the inverted data with a parity word attached. The parallel encoders complete their function within one registered stage, whereas the pipelined encoders have several registered stages to complete their function, so fewer tasks are performed per stage. The two options were explored because it was unknown which architecture would produce a better synthesized result in terms of speed, power consumption, and area. As will be shown in later sections, it became clear that the parallel architecture was a better option so it became the main architecture, as reflected in Table 3. In this section only, the use of the word balanced will also imply nearly-balanced.

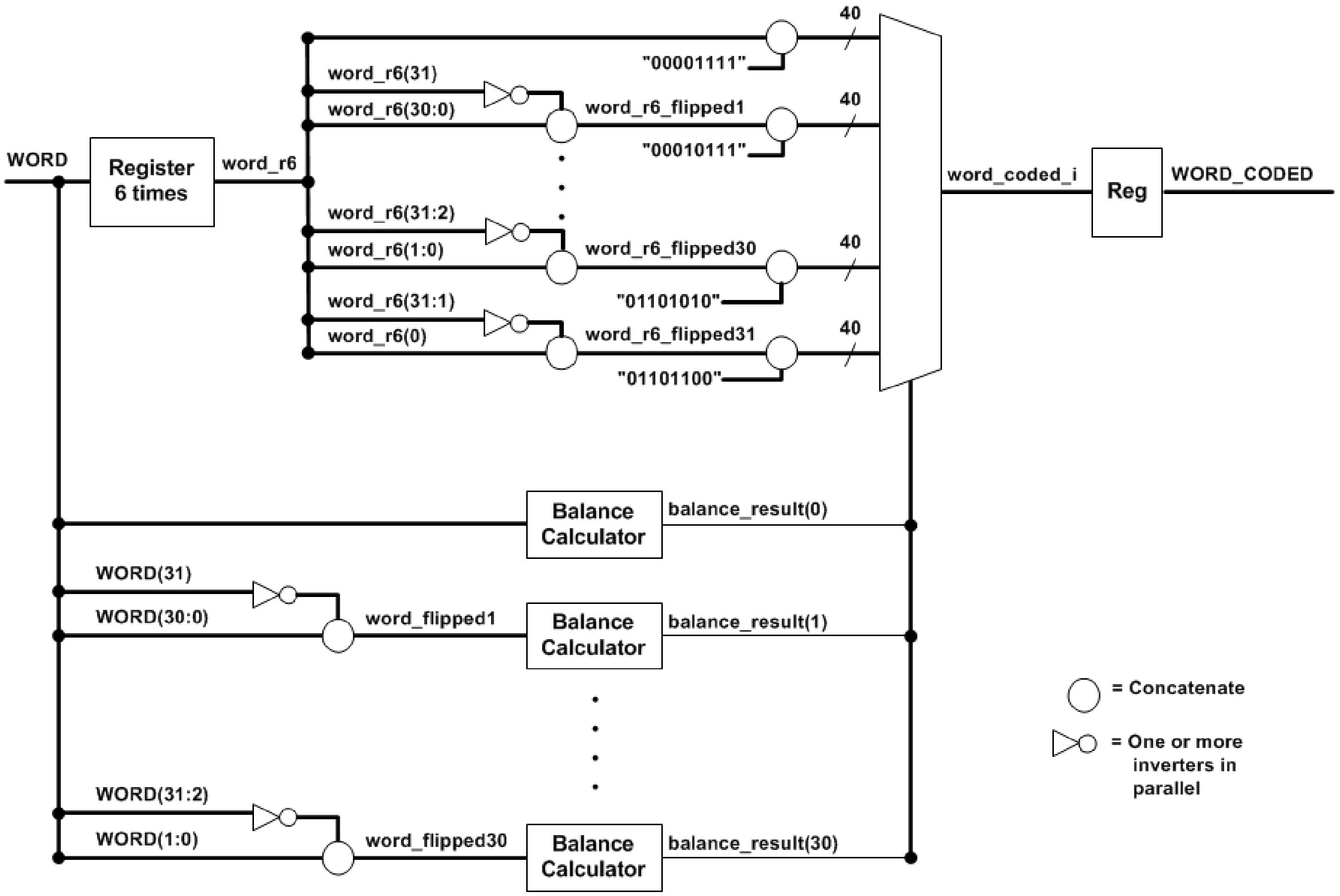

**Figure 9: Example Block Diagram of Encoder Implemented in a Parallel Architecture. Number of Data Bits = 32, Algorithm = Knuth Simple Parallel, Disparity = ±0 (42543v2).**

Figure 9 is a block diagram of a 32-bit parallel architecture where 31 balance calculators simultaneously check disparity on all possible codewords. A balance calculator computes the number of ones in its input using an algorithm from [8], then determines if its input is balanced based on the calculation. In the figure, 0 to 30 bits of the input (WORD) are inverted, starting with the Most Significant Bit (MSB), and then fed simultaneously into the balance calculator blocks. WORD is also registered 6 times to match the latency of the balance calculators then, starting with the MSB, 0 to 30 bits of the registered WORD are inverted. The parity word representing the number of bits inverted in the data is concatenated to the data. The resulting 40-bit coded WORDs are input into the Multiplexer (MUX). More than one balance calculator can have an output of ‘1,’ however the MUX choses the WORD with the least number of inverted bits.

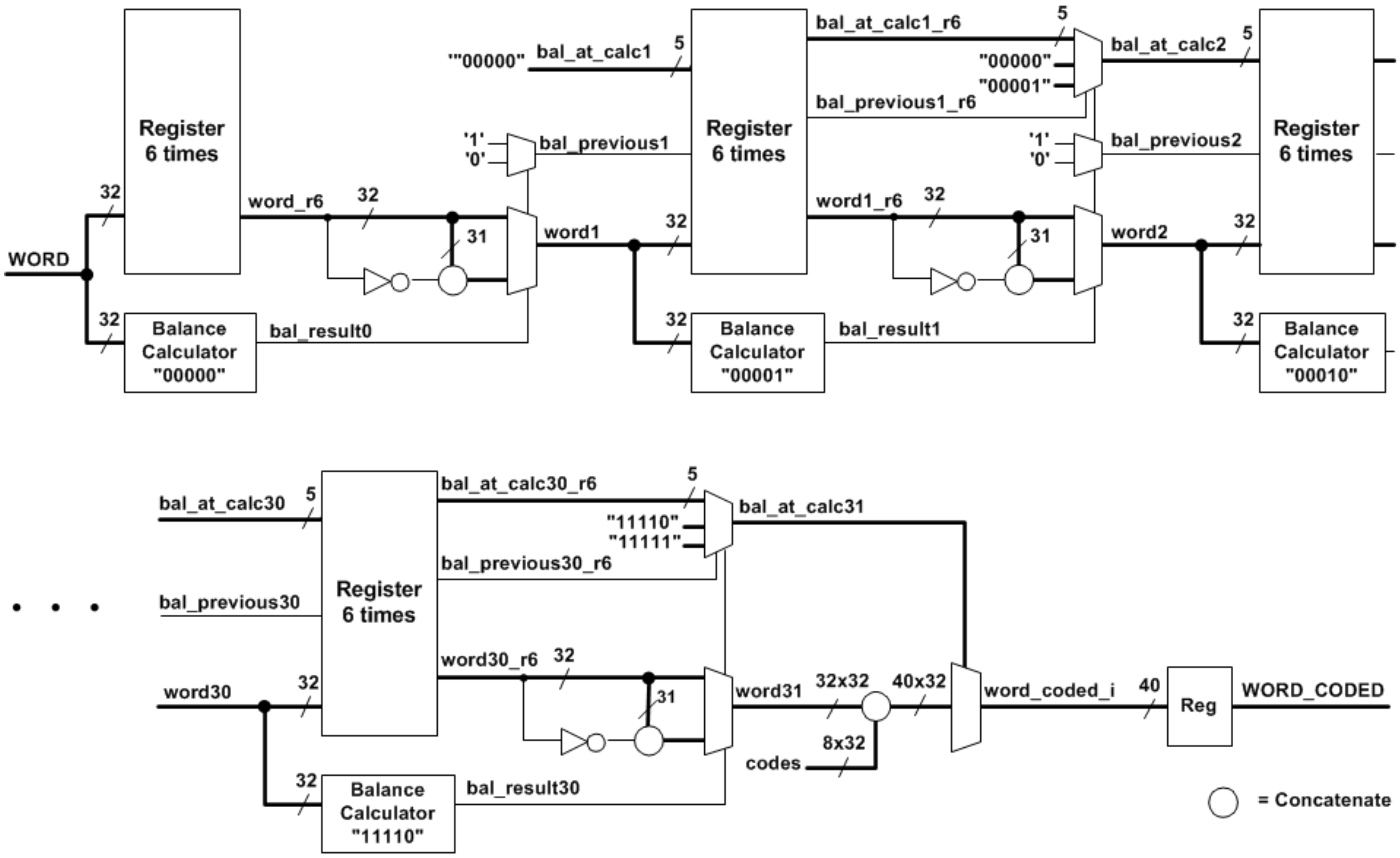

**Figure 10: Example Block Diagram of Encoder Implemented in a Pipeline Architecture. Number of Data Bits = 32, Algorithm = Knuth Simple Parallel, Disparity = ±0 (42542v2).**

Figure 10 is a block diagram of a 32-bit pipeline architecture where balance calculators work independently to find the balance result of data that changes as it moves through the pipeline stages. Data bits in the input (WORD) are fed into the first balance calculator and the first set of registers. The balance calculator result sends either non-inverted data or data with one bit inverted to the next stage. Unlike the parallel architecture, more information has to be sent to the next stages along with the data. Downstream stages need to know if the data was balanced at a previous stage and at what stage that occurred. If the data was previously balanced, downstream stages just pass the data through along with the information about the data. If the data is still unbalanced, a single bit continues to be inverted until the data is balanced. If the data is not balanced after the last stage, then inverting one more bit, which brings the total number of inverted bits in the original data to 31, must produce a balanced result. Therefore, it is not necessary to check the balance of that data. The final data (word31) is replicated 32 times and 32 parity words are concatenated to the appropriate data. The information about which stage produced a balanced result selects which coded word is output.

The performance of both the parallel and pipeline architectures are affected by *n*, the type of Knuth algorithm implemented, and the disparity. In the parallel architecture an increase in *n* increases the number of balance calculators, registers, parity bits, and also increases the size of the MUX. In the pipeline architecture an increase in n will add more stages to the design resulting in an increase in the number of balance calculators and registers. The number of coding bits is also increased. In both the parallel and pipelined architectures, an increase in n increases latency of the balance calculators because more bits need to be counted which increases the number of summation stages in the calculator, so the data must be registered more times in the non-balance-calculator paths in order to match delays. A disparity of ±2 or ±4 decreases the number of balance calculators needed for the design because not all of the *n* choices of bit inversions are used.

The 32-bit designs in Figure 9 and Figure 10 show that both architectures require the same number of balance calculators, 31, but the pipeline architecture requires more registers and more MUXs, therefore the pipeline architecture requires significantly more resources (area, power). Synthesis data presented in the "Encoder Synthesis Results" section will confirm the selection of the parallel architecture as the main architecture for this study.

### Decoder Architecture

The behavioral HDL for the decoders was written simply as case statements embedded within if-else statements. The code was separated from the codeword and several bits of the code were matched to the appropriate if-else condition. Then the remaining code bits were matched to the appropriate case where the appropriate number of bits in the data word were flipped, thus, producing the decoded data word. The synthesizer optimized the behavioral code, dictating the architecture of the decoders.

## Encoder Synthesis Results

Table 3 lists the twenty encoder configurations that were synthesized for this study and the synthesized results for those encoders are shown in Figure 11. The *x*-axis, number of input bits, is the same as *n* used in the "Knuth-Based Algorithms for Balanced and Nearly-Balanced Encoding / Decoding" section.

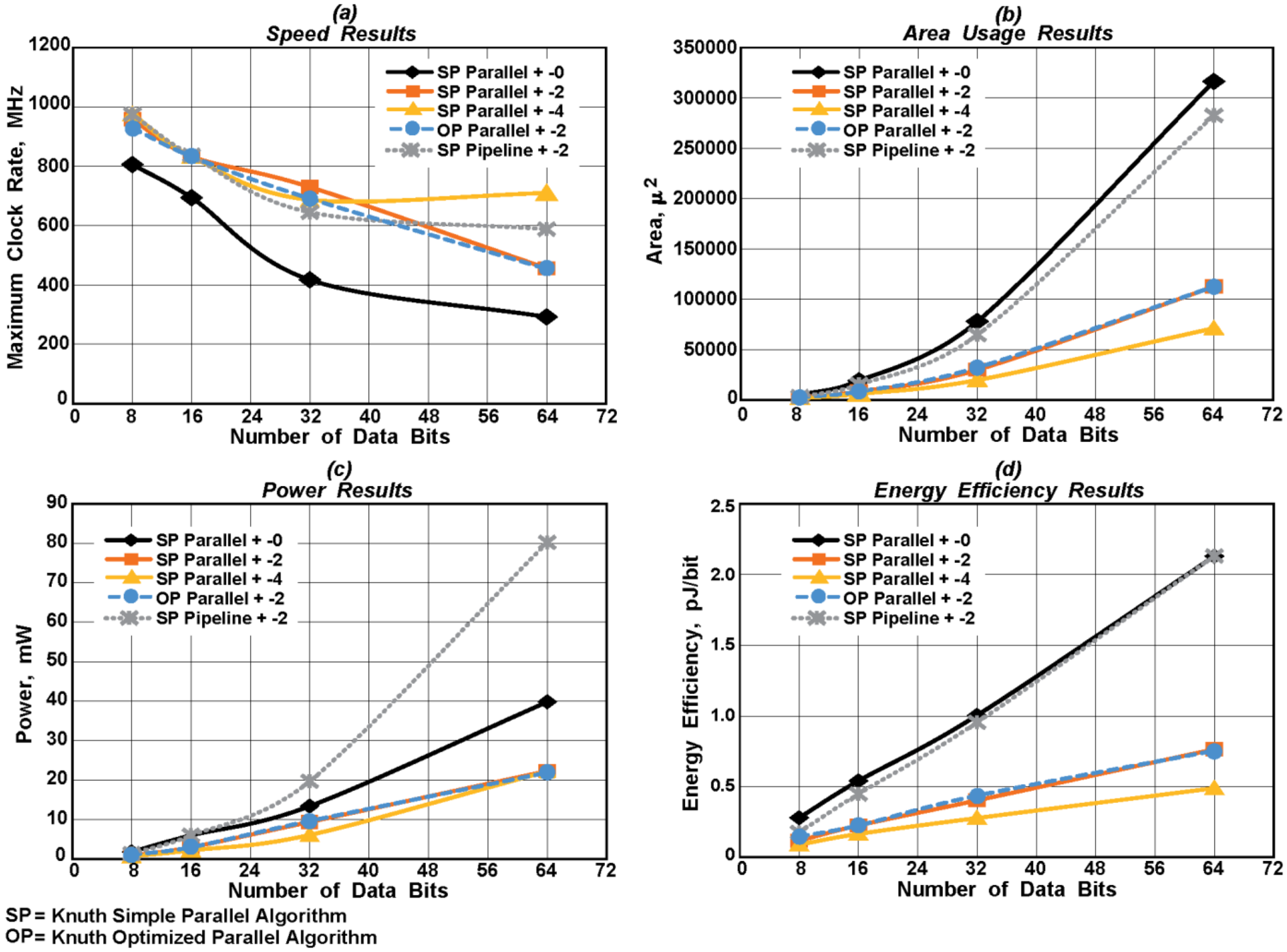

**Figure 11: Encoder Synthesis Results (42550v3).**

Table 4 summarizes the number of logic cells used for each design, which we will refer to as a guide in assessing the size and complexity of each design. As mentioned in the "Parallel and Pipelined Encoder Architectures" section, a lower disparity (i.e. ±2 or ±4) significantly decreases the size and complexity of the designs.

| No. of Input Bits | Design | No. of Cells |
|---|---|---|
| 8 | SP Parallel ±0 | 441 |
| 8 | SP Parallel ±2 | 201 |
| 8 | SP Parallel ±4 | 143 |
| 8 | OP Parallel ±2 | 242 |
| 8 | SP Pipeline ±2 | 300 |
| 16 | SP Parallel ±0 | 1573 |
| 16 | SP Parallel ±2 | 786 |
| 16 | SP Parallel ±4 | 561 |
| 16 | OP Parallel ±2 | 786 |
| 16 | SP Pipeline ±2 | 1250 |
| 32 | SP Parallel ±0 | 6474 |
| 32 | SP Parallel ±2 | 2746 |
| 32 | SP Parallel ±4 | 1771 |
| 32 | OP Parallel ±2 | 2952 |
| 32 | SP Pipeline ±2 | 5287 |
| 64 | SP Parallel ±0 | 26403 |
| 64 | SP Parallel ±2 | 9367 |
| 64 | SP Parallel ±4 | 6399 |
| 64 | OP Parallel ±2 | 9369 |
| 64 | SP Pipeline ±2 | 22312 |

**Table 4: Number of Cells Required for Encoders.**

## Encoder Speed Results

At a high level, several general speed trends are observed:

- Speed decreases sub-proportionally with *n* (i.e. a 64-bit design does not run 8 times slower than an 8-bit design)
- Designs with disparity smaller than zero synthesize to a faster speed than the "SP Parallel ±0" design
- Variations in algorithm (SP vs. OP) have little effect on speed, but the small speed differences which are evident increase with increasing *n*
- Pipelined and parallel architectures operate at comparable speeds, with a slight edge to the pipelined architecture at the highest number of data bits

In general, across a wide range of design parameters, clock rates of greater than 400 MHz are feasible even for modest sized words (64-bit), with 600+ MHz rates demonstrated for smaller words.

## Encoder Area Usage Results

The area usage chart, Figure 11(b), shows the amount of area needed for the synthesized designs in square microns ($\mu m^2$). In general, the complexity (and hence the area) of the designs increases roughly quadratically with increased *n*, as the number of balance calculators increases linearly with *n*, and the size/complexity of each balance calculator also increases roughly linearly with n. SP and OP designs are very similar, using nearly the same number of cells, and hence nearly the same amount of area. As commented above, introduction of finite disparity, *d*, reduces the number of balance calculators needed by a factor of roughly 2*d*+1, with a corresponding reduction in number of cells and area. As the pipelined architecture uses significantly more cells than the parallel (all other things equal), it consumes significantly more area, while operating at comparable speed. In general, all designs are relatively small relative to the size of a typical ASIC input/output cell. Even in the worst case ("SP Parallel ±0", 64 bit), the area consumed is only about 570 μm × 570 μm, or about 70 μm × 70 μm per data bit. In the case of the much more efficient "SP Parallel ±4" design, even in the (worst) 64-bit case, the area per data bit is only about 33 μm × 33 μm.

### Encoder Power and Energy Efficiency Results

Roughly speaking, the power trends, Figure 11(c), can be understood by considering power consumed as proportional to both area (as a doubling in number of cells and hence area roughly corresponds to a doubling of power) and speed (a doubling in clock rate roughly doubles power consumption). Therefore, in general, power increases roughly quadratically with *n* for the same reason that area increases quadratically, but the increase is a bit less pronounced because of the drop in speed with increased *n*. As with the speed and area, power decreases with increasing disparity, and is relatively insensitive to the choice of OP vs SP architecture. The pipelined architecture consumes comparatively more power than the parallel, as it requires significantly more cells.

The energy efficiency results, Figure 11(d), are displayed in pJ/bit, which is a common metric used in assessing efficiency of high speed interconnects. By presenting the data in this way, we provide a simple assessment of how much "overhead" the coding schemes require relative to the rest of the communications circuitry. As a reference point, typical state-of-the-art long-reach SerDes links consume on the order of 10 pJ/bit [9], [10].

Note that the pJ/bit metric is computed by simply dividing the power by both speed and *n*. As commented above, area is roughly proportional to $n^2$, and power is roughly proportional to area times speed, so it is not surprising that the energy consumption per bit increases roughly proportional to *n*. Other trends are as discussed above, explained by considering the design complexity (primarily number of cells).

At the most aggressive *n* evaluated (64-bits), the "worst case" "SP Parallel ±0" and "SP Pipeline ±2" designs consume less than 2.2 pJ/bit. However, the "best case" SP Parallel ±4 design, operates at only 0.5 pJ/bit, with the Parallel ±2 designs only slightly higher (0.75 pJ/bit).

## Decoder Synthesis Results

Synthesized results of the decoders, listed in Table 3, are shown in Figure 12. Details of each metric are discussed in the following sections.

Table 5 is provided to show the number of cells required for each decoder as a guide to assessing design complexity.

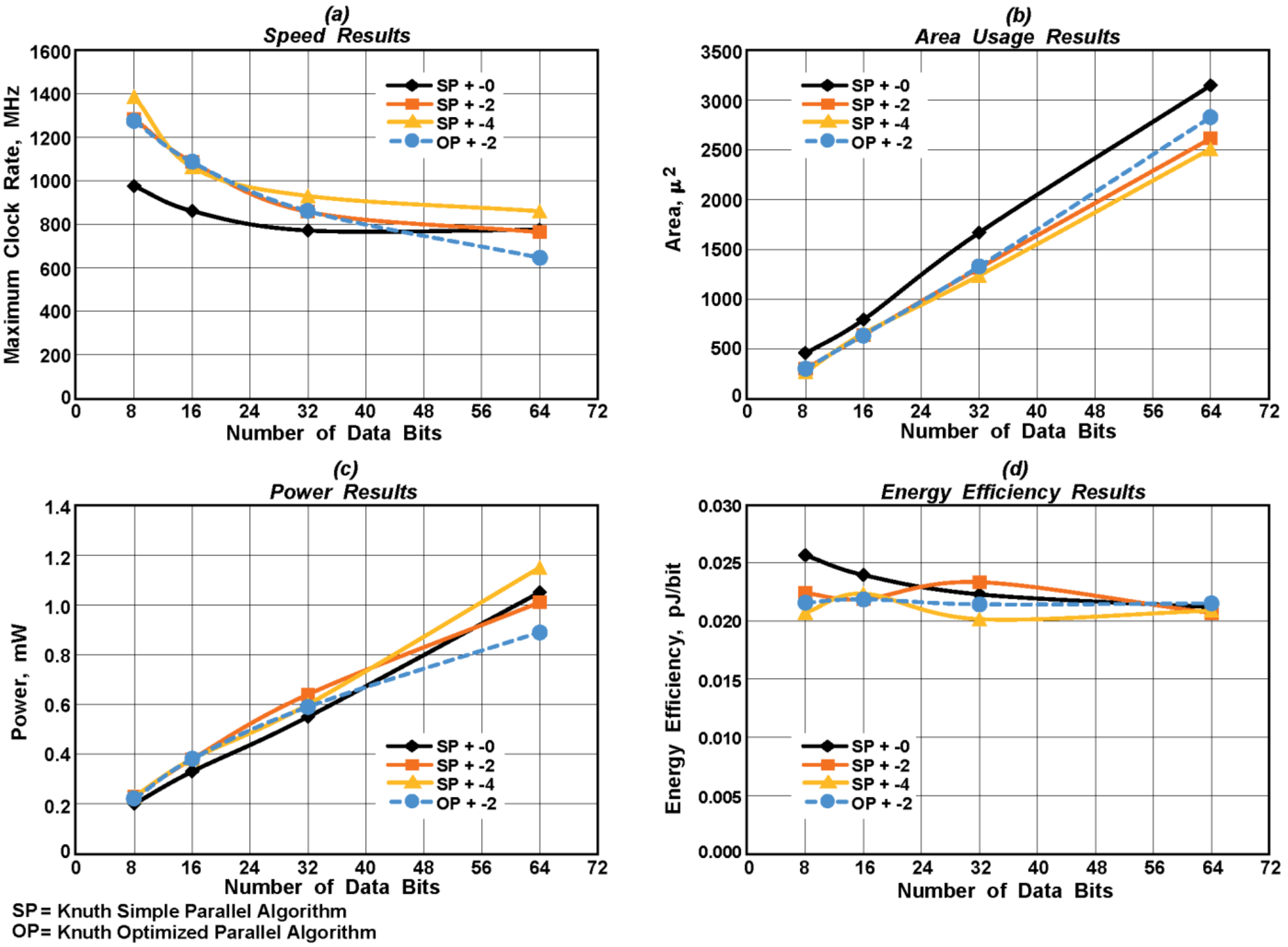

**Figure 12: Decoder Synthesis Results (42551v3).**

| No. of Input Bits | Design | No. of Cells |
|---|---|---|
| 8 | SP ±0 | 47 |
| 8 | SP ±2 | 24 |
| 8 | SP ±4 | 21 |
| 8 | OP ±2 | 25 |
| 16 | SP ±0 | 80 |
| 16 | SP ±2 | 56 |
| 16 | SP ±4 | 57 |
| 16 | OP ±2 | 56 |
| 32 | SP ±0 | 185 |
| 32 | SP ±2 | 119 |
| 32 | SP ±4 | 108 |
| 32 | OP ±2 | 124 |
| 64 | SP ±0 | 336 |
| 64 | SP ±2 | 243 |
| 64 | SP ±4 | 221 |
| 64 | OP ±2 | 276 |

**Table 5: Number of Cells Required for Decoders**

### Decoder Speed Results

The decoder speed results, Figure 12(a), have a similar trend as the encoder results:

- Speed decreases sub-proportionally as *n* increases,
- Designs with disparity smaller than zero generally synthesize to a faster speed than the "SP Parallel ±0" design, and
- Variations in other design parameters have little effect on speed, but these effects tend to increase with *n*.

Decoder speeds can reach above 1 GHz at 16-bits and lower and range from 600 MHz to 900 MHz at 32-bits and higher.

### Decoder Area Usage Results

The area usage results, Figure 12(b), are consistent with the number of cells listed in Table 5.

Note the decoder area increases roughly linearly with *n*, in contrast, the encoder area which increases quadratically. This occurs because the decoders require no repeated block analogous to the encoder balance calculators; rather the decoding "lookup table" grows linearly with *n*. The "SP ±0" designs require the most area while the "SP ±4" designs require the least amount of area above 32-bits, though the differences are minor. Note that area per bit for the decoders (e.g., ~7 µm × 7 µm for the worst case 64-bit "SP ±0" design) is tiny compared to that of the encoder.

### Decoder Power and Energy Efficiency Results

The power results for the Knuth decoders, Figure 12(c), are roughly proportional to area and speed, as discussed above for the encoders, and hence is nearly linear in *n*.

The energy efficiency results, Figure 12(d), are shown in pJ/bit. Somewhat surprisingly, the results across all designs vary at most by 0.006 pJ/bit, (or about 30%) from about 0.020 to 0.026 pJ/bit. With an 8-bit or 16-bit input length the "SP ±0" design is less efficient than all other designs, while all other designs have nearly the same efficiency. At 32-bits the "SP ±4" design is the most efficient but by only a slim (0.0025 pJ/bit) margin. At 64-bits, the efficiency of all designs is basically the same.

## Conclusions

We have assessed the feasibility of implementing fast, area-efficient, and power-efficient balanced and nearly-balanced coding in a 65 nm CMOS technology. In addition to study of two of Knuth's original balanced coding algorithms (Simple Parallel and Optimized Parallel), we demonstrated that these algorithms can be very easily extended to accommodate finite disparity, which can be useful in some applications.

Twenty encoder designs and sixteen decoder designs were synthesized for this study. The designs differed in number of data bits, type of algorithm, disparity and (for encoders only) design architecture. Several general trends were observed:

- Encoders
  - Speed decreases somewhat (sub-proportionally) with increasing *n*
  - Area and power increase roughly quadratically with *n*
  - Energy (pJ/bit) increases roughly linearly with *n*
  - The Parallel architecture outperforms Pipelined by all metrics, except for speed in some cases, as it requires many fewer cells
  - Generally, larger disparity results in simpler designs which improve all metrics, though in some cases by small margins
  - SP and OP algorithms were similar with respect to all metrics

- Decoders
  - Speed decreases modestly with increasing *n*
  - Area and power increase roughly linearly with *n*
  - Energy (pJ/bit) is roughly constant with *n*, and roughly constant across algorithms and disparity
  - Generally, larger disparity results in simpler designs which improve all metrics, though by quite small margins
  - SP and OP algorithms were similar with respect to all metrics

In absolute terms, we have demonstrated the feasibility of encoders in the target 65 nm CMOS technology across a wide range of design parameters, operating at more than 400 MHz clock rate, requiring less than 70 µm × 70 µm area per data bit, and consuming less than 2.1 pJ/bit. Many design points, particularly those using nearly-balanced encoding, significantly exceed these specifications (e.g., the 64-bit "SP Parallel ±4" encoder runs at 700 MHz, requires 33 µm × 33 µm area per bit, and consumes 0.5 pJ/bit). The decoders are significantly simpler circuits than the encoders, and hence run much faster (all designs >600 MHz, most >800 MHz), require much less area (<7 µm × 7 µm per data bit), and consume much less power (<0.026 pJ/bit).

We believe there are many opportunities to further optimize the designs for increased speed, reduced area, and reduced power (and hence improved energy efficiency). Although some investigation of different architectures was considered in this work (e.g., parallel vs. pipelined encoding), we believe there are many opportunities to further improve the encoder architecture in particular (e.g., through more efficient balance calculation algorithms, such as the use of a bitonic sort [11]). In addition, natural benefits would be expected when migrating to more advanced technology nodes (e.g., 32 nm SOI vs 65 nm CMOS bulk).

## Acknowledgments

The authors thank Bill Goetzinger for his help writing the VHDL code and for help running the simulation tools. Thanks are also extended to Terri Funk and Deanna Jensen for their preparation of artwork.